\documentclass[twocolumn,showpacs,amsmath,amssymb,aps]{revtex4-1}
\usepackage{graphicx}% Include figure files
\usepackage{bm}% bold math
\usepackage{dcolumn} % Align table columns on decimal point
\usepackage{subfigure}
\usepackage{color}
\usepackage{epsfig}

\begin{document}
\title{Scanning probe microscopy imaging of metallic nanocontacts}

\author{D. St\"{o}ffler$^{1}$, S. Fostner$^{2}$, P. Gr\"utter$^{2}$ and R. Hoffmann-Vogel$^{1}$}

\affiliation{$^1$Physikalisches Institut and DFG-Center for Functional Nanostructures (CFN), Karlsruhe Institute of Technology, Campus South, D-76128 Karlsruhe, Germany\\
$^2$Department of Physics, McGill University, Montr\'{e}al, Canada}

\date{\today}

\begin{abstract}
{We show scanning probe microscopy measurements of metallic nanocontacts between controlled electromigration cycles. The nanowires used for the thinning process are fabricated by shadow evaporation. The highest resolution obtained using scanning force microscopy is about 3 nm. During the first few electromigration cycles the overall slit structure of the nanocontact is formed. The slit first passes along grain boundaries and then at a later stage vertically splits grains in the course of consuming them. We find that first the whole wire is heated and later during the thinning process as the slit forms the current runs over several smaller contacts which needs less power.}
\end{abstract}

\maketitle

Molecular electronics has been proposed in order to overcome the size limitations of current silicon-based computer technology \cite{aviram74p1}. To investigate possible transistor functionality of single molecules, it is necessary to fabricate three contacts to the molecules of appropriate size \cite{park02,gao09}. Different approaches to obtain nanometer-sized gaps in metallic wires are known, like the mechanically controllable break junction technique \cite{muller92p1,ruitenbeek96p1} or electromigration (EM) \cite{park99p1}. The latter approach offers several advantages. One is that it can be used for many junctions fabricated on the same chip as is needed for applications in computer chips. A second is that it potentially produces gaps accessible to scanning probe tips which can be used for structural characterization of the junction as well as a third moveable gate contact. Coulomb-blockade behaviour with the substrate as a gate has been observed in EM-fabricated nanocontacts \cite{houck05p1}. It can also be observed with scanning force microscopy as has been shown for semiconductor quantum dots \cite{cockins10p1}. The gate efficiency for electromigrated nanocontacts has been investigated by simulations \cite{datta09p1}.

The EM thinning process can be made more reliable either by using a four-terminal configuration \cite{wu07p1}, by making the contacts short \cite{trouwborst06p1} or by controlling the temperature of the junction \cite{strachan05p1,esen05p1}. The electronic properties of the junction during controlled EM have been investigated in air \cite{strachan05p1,hoffmann08}. The junction's structural properties have so far been studied with scanning electron microscopy (SEM) and transmission electron microscopy (TEM) techniques \cite{strachan06p1,heersche07p1,strachan08p1}. In order to run a current through the nanostructure, an insulating substrate is required which complicates scanning tunneling microscopy (STM) studies. Using scanning force microscopy (SFM) we have investigated gaps in small metallic bridges fabricated by controlled EM in UHV. The slit that is formed passes first along grain boundaries. Later the slit crosses through grains. An analysis of the current-voltage characteristics as a function of time allows us to determine that during the process of slit formation, several smaller contacts are formed and that these use less power for EM thinning than is needed at the start of the process.

\begin{figure}
        \begin{center}
	    \includegraphics[width=\linewidth,angle=0,clip]{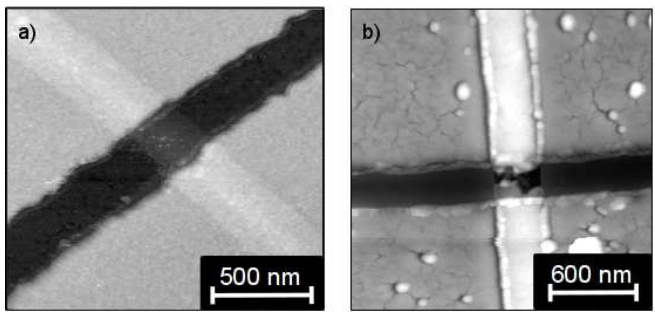}
 	        \end{center} 
	\caption{
a) STM image of an electron-beam evaporated Au wire of 7 nm thickness (U$_T$ = 10V, I$_T$ = 0.5 nA, $\Delta$z = 37 nm) and b) AM-SFM image of a sputtered Pd wire after EM thinning ($\Delta$z = 233 nm)}
	\label{stoe1_f1}
\end{figure}

\begin{figure}
        \begin{center}
	    \includegraphics[width=\linewidth,angle=0,clip]{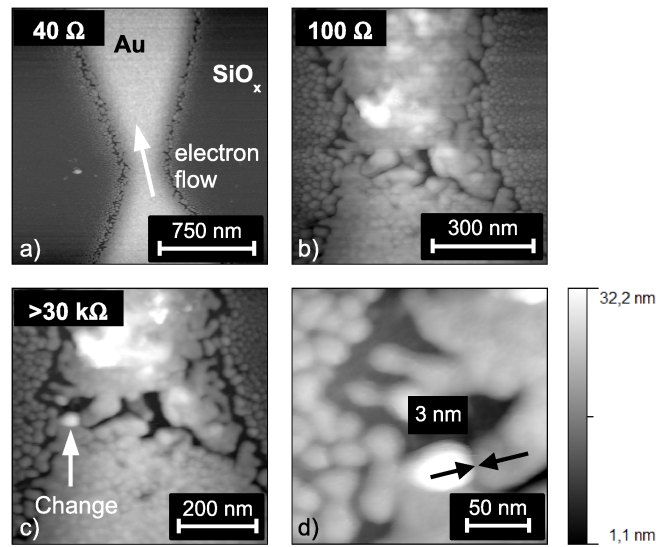}
 	        \end{center}
	\caption{AM-SFM images of Au nanowire of 30 nm nominal thickness before (a) and after (b) EM. The resolution limit is around 3 nm. The initial resistance of the wire was 40 $\Omega$ and the EM started at around 1.2 V.}
	\label{stoe1_f2}
\end{figure}

\begin{figure}
        \begin{center}
	    \includegraphics[width=0.7\linewidth,angle=0,clip]{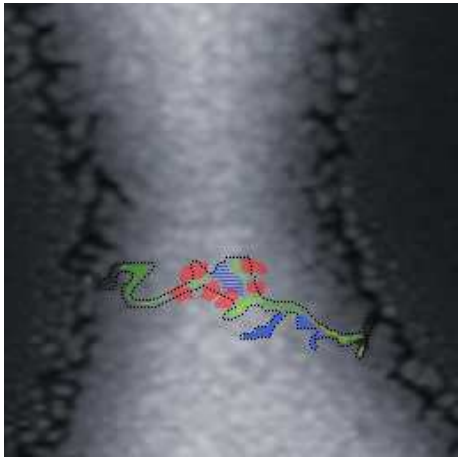}
        \end{center}
	\caption{(Colour online) Colour-coded results of a careful analysis overlayed with an SFM image of the original nanowire (Fig. 2). Positions where the slit passes along grain boundaries - green - where the slit has removed grains completely - blue. Hatched blue regions are partly removed grains. Grains through which the slit has passed - red.}
	\label{stoe1_f3}
\end{figure}

We fabricated metallic nanobridges (Pd, Au) either by standard e-beam lithography (Fig. \ref{stoe1_f1}) or by shadow evaporation in ultrahigh vacuum (Fig. \ref{stoe1_f2} and Fig. \ref{stoe1_f3}). For the latter technique, mask manipulating devices that can be operated in UHV similar to the ones shown in \cite{gaertner06p1} are used to bring a Si or Si$_{3+x}$N$_{4-x}$ membrane mask nanostructured by focussed ion beam to close proximity of an oxidised Si substrate. The metallic layers of nominal thickness 20-100nm were deposited either by e-beam evaporation or by sputtering. For samples fabricated by e-beam lithography, we added the macroscopic leads in a second step while for the ones prepared by shadow evaporation, the complete structure was deposited in one evaporation step. We imaged the structures by SEM, UHV-STM (Omicron with Nanonis controller) and UHV-SFM (JEOL with Nanosurf Phase Locked Loop). For the STM measurements, we prepared tungsten tips by electrochemical etching, in-situ sputtering at 2.5 keV for 10 min, additional e-gun heating at 1 mA for 15 min and again sputtering for 5 min. For the SFM measurements, we used commercially available silicon cantilevers with a resonance frequency of about 170 kHz, a typical $Q$-factor of 5 000 and a spring constant of 40-50 N/m.

We were able to image the nanostructure by STM inspite of the insulating oxide layer covering the Si substrate using tunneling voltages between 8 and 10 V (Fig. \ref{stoe1_f1}a)). However, these imaging conditions are not ideal for obtaining high resolution. We therefore moved to SFM imaging. In addition, on e-beam lithography made samples, a contamination layer was formed on the sample in STM images. The layer is caused through the contact with chemicals during electron beam lithography fabrication and immobilization through the flow of electrons during STM imaging \cite{stoeffler09p1}. We therefore moved to shadow evaporation to provide clean surfaces necessary for high-resolution measurements.

For the SFM images (Fig. \ref{stoe1_f1}b), Fig. \ref{stoe1_f2} and \ref{stoe1_f3}) we used amplitude-modulation mode (AM-SFM) mode, where the cantilever is oscillated at a constant frequency off-resonance and the reduction of the amplitude upon approach is used as a measure of the tip-sample interaction. This became possible by reducing the $Q$-factor of the setup with $Q$-control technique \cite{holscher06p1}. For these three-dimensional samples we obtained better results than with frequency-modulation mode \cite{moritab} that is usually used under vacuum conditions.

The overall microstructure of the metallic layers depends on the material (Au, Pd, Pt) and on the deposition method and is well-known from previous studies. Sputtered Pd layers show smoother surfaces than e-beam deposited Au layers. For e-beam lithography-made samples, often a topographically elevated rim near the edge of the PMMA mask is observed that can be explained by the carpet growth model for metallic thin films. For samples fabricated by shadow evaporation a half-shadow region \cite{gaertner06p1} is observed in which the total nominal film thickness is reduced compared to the center of the nanostructures and in which electrically disconnected metallic grains occur (Fig. \ref{stoe1_f2} and Fig. \ref{stoe1_f3}).

The electrical characterization of the samples showed that sputtered Pd samples have higher overall initial resistances (R$_0$=400-650 $\Omega$) compared to the Au samples prepared by electron-beam evaporation (R$_0$=40 $\Omega$). Both resistances are significantly increased compared to the values expected using the bulk resistivity and the dimensions of the wire, about 1 $\Omega$ for Au and 4 $\Omega$ for Pd, due to the microstructure of the wires and possible contact resistances. It is well-known that the resistances of sputtered wires are usually higher than the resistances of electron-beam evaporated wires. After an initial image we thinned the nanobridge by controlled EM by applying a computer-controlled voltage to the samples in cycles similar to the procedure described in \cite{esen05p1,strachan05p1,hoffmann08}. For each cycle, we increased the voltage in steps of 1 mV per 100 ms until the initial resistance increased by  1-6$\%$. Then the voltage was automatically reduced so that the power was decreased to 20$\%$ and a new cyle was started \cite{hoffmann08}. We stopped the thinning process at suitable pre-chosen resistances to take additional images.

\begin{figure}
        \begin{center}
	    \includegraphics[width=1.0\linewidth,angle=0,clip]{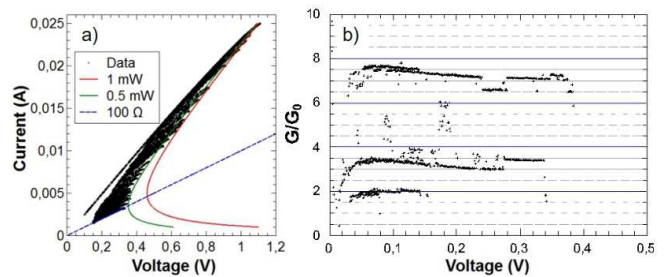}
 	        \end{center}
	\caption{(Colour online) a) Overall EM process with two lines of constant dissipated power as a guide to the eye. b) Conductance of the nanocontact below ten conductance quanta observed during EM thinning.}
	\label{stoe1_f4}
\end{figure}

After 2-8 controlled EM cyles on all samples a slit had formed. One particularly complete measurement is shown in Fig. \ref{stoe1_f2} to \ref{stoe1_f4}. This particular sample had been prepared by shadow evaporation of 30 nm Au onto a Si substrate covered by native oxide through a nanostructured mask. It had an initial resistance of 40 $\Omega$. We took a SFM image at a total resistance of 100 $\Omega$. The slit typically has a width of 10 - 20 nm, although at particular positions it is as wide as 80 nm. Hillocks have build up following the direction of the electron flow from the slit to the top electrode. These hillocks are expected to build up through thermally assisted diffusion of metal under the influence of EM forces, i.e. the wind force caused by momentum transfer from the flow of electrons and the electrostatic forces. On the slit one can see several positions where the sample may still be connected or a small gap may have formed. The exact position of the point where the contact is still connected has been determined only after electrically disconnecting the contact (see below) and is marked by an arrow in Fig. \ref{stoe1_f2} c).  We furthermore stopped the EM to take SFM images at 167, 328, 630 and 12 900  $\Omega$. These images showed no further apparent structural change of the contact due to the resolution of the SFM images. We evaluated the AFM imaging resolution as the distance marked by black arrows in Fig. \ref{stoe1_f2}d) on an even smaller image and estimate the best resolution to be about 3 nm.

Since no systematics in the different shapes of the nanocontacts in our own work nor in literature is observed the overall structure of the contact after significant EM must be a direct result of the EM process itself. The initial stages of the EM process have been studied previously \cite{holm67}. At the start of the process, the contact is in the thermal regime, where the voltage and the temperature have a fixed relation. Since the microstucture of the wire includes defects like impurities and grain boundaries (see also \cite{baldini93} for example), there is a distribution of binding energies of the atoms in the wire. As the voltage is increased and thus the temperature of the wire is increased, the atoms with the smallest binding energies first start to hop due to thermal activation. Since the atoms at grain boundaries, in dislocations and at the surface of the wire have smaller binding energies than the bulk, these atoms first start to modify their positions.

In the region where material has been removed, the temperature must have been locally larger than the temperature needed for diffusion while at the same time at the remaining parts of the contact the local temperature was below that temperature. For an analysis we carefully overlayed the SFM images obtained before and after EM (Fig. \ref{stoe1_f3}, movie available online \cite{EPAPS}). In Fig. \ref{stoe1_f3}, the region where material has been removed is surrounded by two dotted lines. The movie allowed us to identify several parts of the slit: in some regions, the grains originally present in the Au film fully disappeared as the slit formed (blue regions) while in other regions the slit followed the grain boundaries (green). Some grains were split vertically by the slit formation (red). In the hatched blue regions, grains were removed partly and split horizontally. There is a tendancy to find the red grains in the center of the nanowire while the blue grains are found on the right hand side and only one hatched blue area is found in the center of the nanowire.

Initially the full width of the connected grains forming the nanowire is rather uniformly heated by the flowing current with a smooth decay towards the leads. Such a uniform temperature distribution is in accordance with the formation of a slit of rather uniform width along the grain boundaries, because the atoms in the grain boundaries are expected to have a similar binding energy, and when they are subject to EM we therefore expect them to become mobile at a common temperature. The green area that is not connected to the slit on the right hand side of the wire shows that at least part of the slit formed by pinching in from the sides as has been observed previously \cite{heersche07p1,strachan08p1}. A uniform temperature distribution can however not explain why single grains are removed (blue grains) or removed partly (red grains). To explain these findings we suppose that the temperature distribution in the nanowire has changed in the course of slit formation as the current density and thus temperature varied locally along the forming slit. For the red grains we infer that the local temperature must have varied \textit{within one grain} while the temperature variation for the blue grains was rather smooth on the length scale of the grains. We associate such a strongly localized temperature variation with the formation of smaller contacts and more localized current paths in the course of slit formation. As a consequence void formation that has affected the blue grains must have occurred first, when the temperature distribution was still broad, while the formation of the slit near the read grains must have occurred later, when the temperature distribution was locally sharp in smaller remaining nanocontacts. The temperature could vary on the length scale of the mean free path of the metal which we expect to be on the order tens of nm for the bulk value, but it could be an order of magnitude smaller than that for our thin film. This should be compared to the typical size of the grains - 25-35 nm. Considering the typical size of the slit of 10 - 20 nm, one should consider that once parts of the slit have formed the shape of the contacts could be further influenced by Fowler-Nordheim tunneling, a process which is known to consume material at the sharpest points of the contacts.

These considerations can be linked to the electronic measurements during EM. Lines of constant dissipated power at the junction correspond to hyperbolic curves \cite{hoffmann08,wu07p1}. The I(V) characteristics  show that for this junction the maximal dissipated power in each oscillation cycle was not constant, but remained at 1 mW up to a resistance of 55 $\Omega$ and then lowered automatically to 0.5 mW. This can be understood by the smaller volume that was heated when the current was running through smaller remaining contacts. We have observed a similar behaviour in other contacts.

We then proceeded with the controlled EM process and reached the ballistic regime (Fig. \ref{stoe1_f4}), where we observed conductance plateaus at not necessarily integer multiples of the conductance quantum $G_0$ and discrete conductance jumps (Fig. \ref{stoe1_f4}) in agreement with previous findings \cite{hoffmann08}. After the last EM-cycle, we subjected the nanocontact to 5 V to make sure it was definitely broken. Another SFM image showed two small spheres (diameter 20 - 50 nm) indicating that the material had molten locally and assumed the shape of a sphere to minimize surface energy (Fig. \ref{stoe1_f2}c) and d)). We conclude that the wire had been connected at the position where the two spheres had formed. The closest final distance between the two electrodes in the region of the spheres was imaged with best possible resolution (Fig. \ref{stoe1_f2}d)) and was found to be similar to $3$ nm, our resolution limit.

Looking at the overal microstructure of the nanocontact after its formation (Fig. \ref{stoe1_f3}) we observe that there has been no significant recrystallization in this particular contact. If such a recrystallization had taken place, we would have expected a difference of the microstructure near the slit compared to positions further away from the slit. This could be explained if the temperature during EM was rather low compared to the relevant activation energy barriers and also compared to other studies (e.g. \cite{gao09}), where significant recrystallization is observed including some of our other samples. Recrystallization could also be enhanced by surfactant effects caused by impurities. This should be checked further.

In conclusion we have studied the structural evolution of nanowires during controlled EM with scanning probe techniques in UHV. Best resolution was obtained in dynamic scanning force microscopy in the amplitude modulation mode. During EM a slit is formed at the center of the nanowire. We identify the grain boundaries and grains through which the slit passes. The comparison of the current-voltage characteristics with the topographic images allows us to identify two main regimes in the thinning process: first the volume of the fabricated nanostructure is heated near its thinnest point and parts of a slit are formed by consuming material at the grain boundaries. Then a smaller volume is heated as the contact breaks up into several smaller contacts and the slit formation progresses by passing through single grains. This process requires less power than was used in the first part.

Financial support from the Baden-W\"urttemberg-Stiftung in the framework of its excellence program for postdoctoral researchers, from the European Research Council through the Starting Grant NANOCONTACTS (No. 239838), from the Karlsruhe House of Young 
Scientists, from NSERC, FQRNT, CIfAR and CFI is gratefully acknowledged.

\end{document}